# Intrinsic exciton transport and recombination in single-crystal lead bromide perovskite

*Zhixuan Bi,*[1] *Yunfei Bai,*[2] *Ying Shi,*[1,3] *Tao Sun,*[4] *Heng Wu,*[5] *Haochen Zhang,*[1] *Yuhang Cui,*[6] *Danlei Zhu,*[6] *Yubin Wang,*[1,3] *Miao-Ling Lin,*[5] *Yaxian Wang,*[2] *Dongxin Ma,*[6,7] *Ping-Heng Tan,*[5] *Sheng Meng,*[2,*] *Qihua Xiong,*[1,3,8,*] *and Luyi Yang*[1,8,*]

[1] State Key Laboratory of Low Dimensional Quantum Physics, Department of Physics, Tsinghua University, Beijing 100084, China

[2] Beijing National Laboratory for Condensed Matter Physics and Institute of Physics, Chinese Academy of Sciences, Beijing, 100190, China

[3] Beijing Academy of Quantum Information Sciences, Beijing 100193, China

[4] Institute of Physics, Chinese Academy of Sciences, Beijing 100190, China

[5] State Key Laboratory of Semiconductor Physics and Chip Technologies, Institute of Semiconductors, Chinese Academy of Sciences, Beijing 100083, China

[6] Key Lab of Organic Optoelectronics and Molecular Engineering of Ministry of Education, Department of Chemistry, Tsinghua University, Beijing 100084, China

[7] State Key Laboratory of Flexible Electronics Technology, Tsinghua University, Beijing 100084, China

[8] Frontier Science Center for Quantum Information, Beijing 100084, China

*Corresponding author. Email: smeng@iphy.ac.cn; qihua_xiong@tsinghua.edu.cn; luyi-yang@mail.tsinghua.edu.cn

ABSTRACT

Photogenerated carrier transport and recombination in metal halide perovskites are critical to device performance. Despite considerable efforts, sample quality issues and measurement techniques have limited the access to their intrinsic physics. Here, by utilizing high-purity $CsPbBr_3$ single crystals and contact-free transient grating spectroscopy, we directly monitor exciton diffusive transport from 26 to 300 K. As the temperature ($T$) increases, the carrier mobility ($\mu$) decreases rapidly below 100 K wtih a $\mu \sim T^{-3.0}$ scaling, and then follows a more gradual $\mu \sim T^{-1.7}$ trend at higher temperatures. First-principles calculations perfectly reproduce this experimental trend and reveal that optical phonon scattering governs carrier mobility shifts over the entire temperature range, with a single longitudinal optical mode dominating room-temperature transport. Time-resolved photoluminescence further identifies a substantial increase in exciton radiative lifetime with temperature, attributed to increased exciton population in momentum-dark states caused by phonon scattering. Our findings unambiguously resolve previous theory-experiment discrepancies, providing benchmarks for future optoelectronic design.

MAIN TEXT

Over the past decade, metal halide perovskites have garnered great attention for their superior optoelectronic properties and high performance in solar cells and light-emitting devices.[1,2] As a photoactive material, the transport and recombination of photogenerated carriers are critical processes that have been extensively studied using various spectroscopic[3–9] and electrical[10–12] techniques, leading to a consensus on their modest mobilities and long carrier lifetimes at room temperature.[13,14] While the impact of extrinsic factors, such as impurities and grain boundaries, can be minimized through defect passivation and improved crystalline quality, intrinsic factors originating from inevitable interactions between carriers and the perovskite lattice fundamentally define the limits on their optoelectronic performance.

Understanding the nature of electron-phonon coupling in metal halide perovskites and its impact on carrier transport and recombination is therefore essential across single crystals, polycrystalline films, and nanocrystals. Studies on the temperature dependence of charge-carrier mobility have consistently observed a scaling behavior of $\mu \sim T^{-1.5}$ near room temperature.[6–12] For a long time, this temperature trend was interpreted as an experimental signature of deformation potential scattering by acoustic phonons[9–11,15] based on a highly simplified model.[16] However, early calculations have shown that acoustic phonon scattering is relatively weak and should result in much higher mobility on the order of several thousand $cm^2V^{-1}s^{-1}$ at room temperarture,[17,18] which is much higher than the observed values. Some articles suggest that large polaron formation, arising from Fröhlich coupling, can explain the modest mobility through an enhanced effective mass.[19,20] However, the polaron model predicts the correct magnitude[21] but fails to account for the observed temperature dependence.[22,23]

Given the soft and ionic nature of the perovskite lattice, the strong Fröhlich interaction between electrons and longitudinal-optical (LO) phonons is instead expected to be the dominant source of electron-phonon coupling, as supported by temperature-dependent analyses of emission line broadening[24] and recent ab initio calculations[25,26]. Nonetheless, polar optical phonon scattering typically results in a more pronounced power-law scaling with temperature,[27] which conflicts with the observed acoustic-like $T^{-1.5}$ behavior. To date, a clear understanding of the transport properties in halide perovskites remains elusive, and the contributions from different phonon modes have yet to be fully delineated.

An important next step requires experimental studies on high-quality single-crystal samples, as emphasized in several theoretical articles.[19,22,23] Moreover, because most experiments on the temperature scaling have been performed in the range between 100 K and room temperature, it is highly desirable to measure the low-temperature regime, where optical phonons are thermally suppressed. However, probing intrinsic transport in perovskite semiconductors at very low temperatures is challenging[8,12] due to difficulties in obtaining ultra-pure samples that can rule out extrinsic effects from impurities and grain boundaries, as well as the lack of reliable measurement techniques free from systematic errors, such as contact resistance. Although single crystals serve as an ideal research platform, the transport properties revealed from them are equally applicable to perovskite polycrystalline films and nanocrystals.

In addition to transport, understanding the recombination properties is also crucial for the application of perovskite-based devices. Early reports have shown extremely long carrier lifetimes and diffusion lengths in solar-cell perovskites,[28] where fast dissociation of bound electron-hole pairs (excitons) enables high charge collection efficiency and slow recombination kinetics. In contrast, all-inorganic $CsPbBr_3$ sustains stable excitons at room temperature[29] and has been the focus of much recent work aiming at achieving efficient photon emission in nanocrystals.[30] In intrinsic semiconductors, the lifetime of radiative recombination generally increases with rising temperature,[31,32] a trend that is rarely observed in bulk perovskites due to the competition from trap-related recombination decay. The greatly enhanced exciton binding energy in nanocrystals, resulting from spatial confinement, makes this behavior more readily observable[30,33], and a similar temperature-dependent trend is expected in high-purity bulk counterparts.

In this work, we investigate the intrinsic exciton transport and recombination mechanisms in high-purity vapor-phase epitaxially grown $CsPbBr_3$ single crystals using transient grating (TG) spectroscopy.[34] This powerful all-optical technique avoids the contact issues inherent in electrical measurements and has been successfully applied to explore photogenerated charge,[35,36] spin,[37,38] and thermal transport.[36,39,40] Our experiments, conducted from room temperature down to ~26 K, reveal a modest exciton diffusion coefficient of 0.58 $cm^2s^{-1}$ at room temperature, corresponding to an effective mobility of 22.8 $cm^2V^{-1}s^{-1}$. The effective mobility increases to ~$10^4$ $cm^2V^{-1}s^{-1}$ at 26 K. The temperature-dependent mobilities exhibit two distinct scaling behaviors, best fitted by $\mu{\sim}T^{-3.0}$ below 100 K and $\mu{\sim}T^{-1.7}$ above 100 K. These temperature trends are fully reproduced by our first-principles calculations, revealing that acoustic phonons play a negligible role in carrier transport across the entire temperature range. The rapid decline in mobility below 100 K is driven by the onset of strong Fröhlich interactions with soft LO phonons, while the acoustic-like scaling above 100 K results from the dominance of LO phonon scattering. Additionally, we observe an increase in exciton lifetime with rising temperature, which can be explained by enhanced exciton scattering of by phonons into optically momentum-dark states, effectively suppressing radiative recombination. The intrinsic temperature dependence of the exciton radiative lifetime, which has most commonly been observed in nanocrystals,[30,33] further highlights the high quality of our epitaxial samples. Our findings not only resolve the ongoing debate on the microscopic mechanisms governing carrier transport in metal halide perovskites but also offer valuable insights for future design of optoelectronic devices based on these materials.

## RESULTS

### Intrinsic exciton states in high-purity $CsPbBr_3$ single crystals

Single crystalline $CsPbBr_3$ perovskite thin films were epitaxially grown on freshly exfoliated muscovite mica substrates using a custom-built chemical vapor deposition system.[41] As shown in Figure 1a, optical microscopy reveals distinct single-crystalline domains with lateral sizes ranging from tens to hundreds of μm. Compared to perovskite single crystals grown by inverse temperature crystallization or high-vacuum evaporation,[42,43] these epitaxial films exhibit superior long-range crystallinity, exceptionally smooth surfaces (with a roughness of less than 0.2 nm) and significantly thinner thickness (several hundred nm, Figure S3). Moreover, these single-crystalline films can be easily exfoliated and transferred onto various substrates, such as sapphire (Figure S2) and silicon wafers, making them an ideal platform for studying the intrinsic properties of halide perovskites using a range of electrical and optical techniques.

Figure 1b presents the powder and single-crystal X-ray diffraction (XRD) patterns of the synthesized single-crystalline films. Note that only the lattice planes parallel to the (001) plane of the mica substrate contribute to the powder XRD patterns. The splitting of the (110) and (002) diffraction peaks, together with the $hk0$ reciprocal lattice plane reconstructed from single crystal

XRD data, clearly reveals that the epitaxial $CsPbBr_3$ films adopt a tetragonal phase at room temperature. Upon cooling, a phase transition from the tetragonal to the orthorhombic phase is observed at ~260 K (Figure S1).

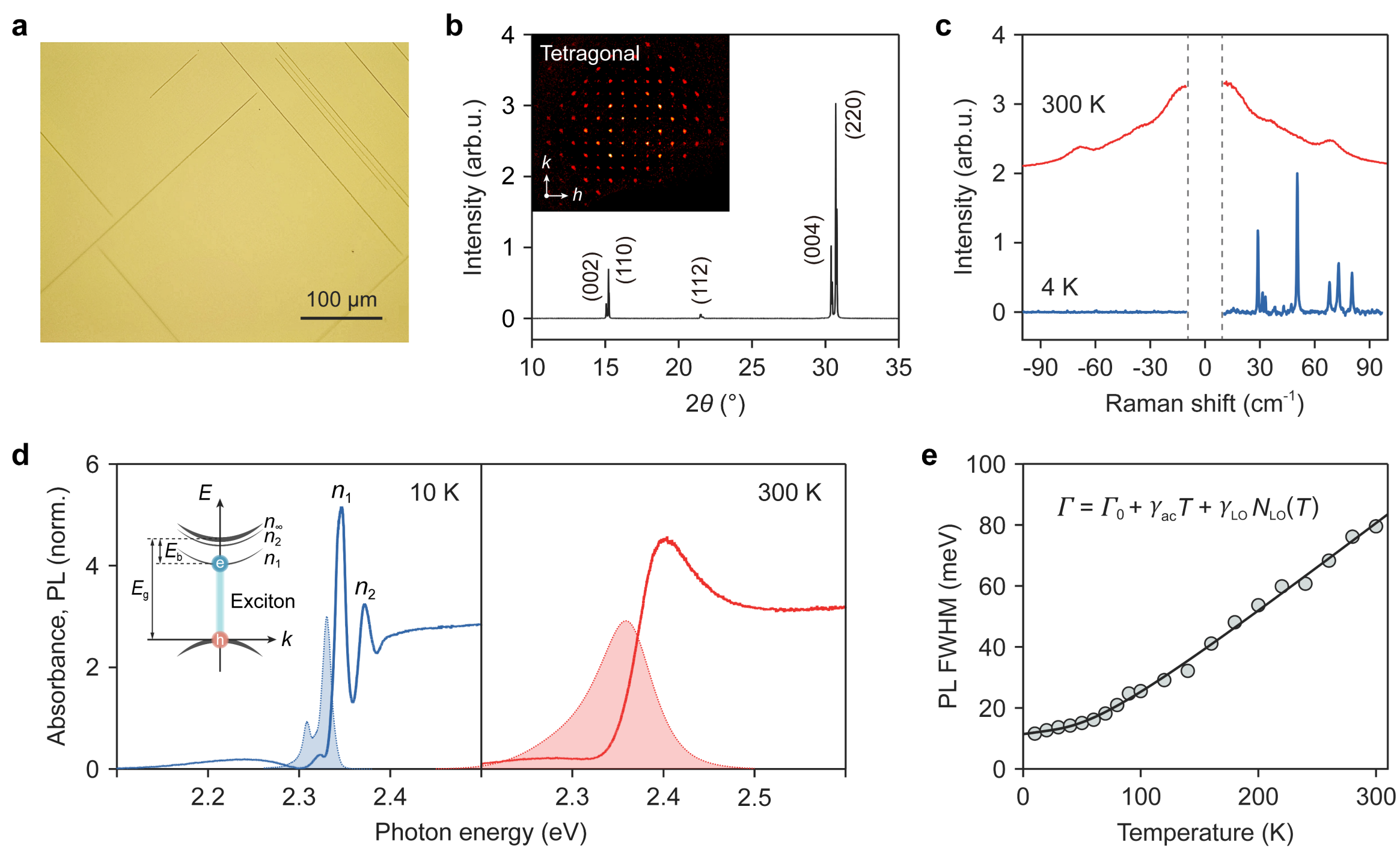

Figure 1. Optical characterization of single-crystalline $CsPbBr_3$ thin film. (a) Optical microscopy image of the epitaxial $CsPbBr_3$ thin film, showing perfectly flat surfaces and relatively large domains for optical measurements. The well-oriented cracks arise from the mismatch in thermal expansion coefficients between the epitaxial perovskite layer and the mica substrate during crystal growth. (b) Room-temperature powder and single-crystal XRD patterns. Detailed structure refinement identifies a tetragonal phase at room temperature. (c) Typical low-frequency Raman spectra measured at cryogenic and room temperatures, revealing a series of optical phonons below 100 $cm^{-1}$. Curves are vertically offset for clarity. (d) Absorbance (solid lines) and normalized PL spectra (shaded areas). (Left) At 10 K, the first- and second-order exciton transitions are clearly resolved in the absorbance due to the extremely sharp linewidth. The inset illustrates multiple Rydberg energy levels. In the PL spectrum, the second peak at a lower energy is attributed to the formation of self-trapped excitons. (Right) At 300 K, the spectra are significantly broadened by thermally activated phonons; however, strong excitonic absorption persists due to the ultrastable excitons in high-purity $CsPbBr_3$. (e) FWHM of the dominant PL peak as a function of temperature (gray dots). The black line represents a phenomenological fit, accounting for scattering from impurities and phonons. arb. u. stands for arbitrary units.

The Raman spectrum at 4 K shows that $CsPbBr_3$ exhibits a series of low-energy optical phonons below 100 $cm^{-1}$ (Figure 1c), highlighting the soft nature of the perovskite lattice. These soft optical modes can be significantly populated by modest thermal excitation, enabling electron-optical phonon interactions to influence carrier dynamics even at relatively low temperatures. In addition, at elevated temperatures, strong central peak emerge due to anharmonic local polar fluctuations.[44]

Figure 1d presents the absorbance and photoluminescence (PL) spectra of our sample at low (10 K) and high (300 K) temperatures. In semiconductors, optical absorption can generate electron-hole pairs (excitons) due to Coulomb interaction, forming hydrogen-like bound states that significantly enhance absorption below the band edge. Applying a generalized hydrogen model for free excitons, the transition energy of the $n$th ($n=1,2,3,\ldots$) exciton state is $E_n = E_g - E_b/n^2$, where $E_g$ is the band gap and $E_b$ is the binding energy. At 10 K, the absorbance spectrum distinctly reveals the $n=1$ and $n=2$ Rydberg exciton states, with higher states merging with the continuum at the band edge. The clear identification of discrete exciton states has also been previously observed in high-quality bulk semiconductors such as GaAs[45] and $Cu_2O$[46]. From these two transitions, we directly extract a band gap of $E_g$=2.38 eV and an exciton binding energy of $E_b$=35 meV, which agrees well with the 33-meV value determined from magneto-transmission experiments.[47] The strong excitonic absorption peak persists at 300 K, indicating a stable exciton population resistant to thermal ionization.

Both the absorbance and PL spectra reveal a monotonically increasing band gap, accompanied by a broadened feature as the temperature rises (Figure S5). Notably, the absorbance, calculated directly from the transmittance, shows an anomalous bump below the absorption onset, likely due to surface reflection effects.[48] In the PL line shape, a second peak at lower energies is often linked to the formation of self-trapped excitons.[49]

Figure 1e shows the full width at half maximum (FWHM) of the dominant PL peak as a function of temperature, providing insights into exciton-phonon coupling mechanisms. The temperature-dependent line broadening is generally analyzed by the expression[24] $\Gamma(T) = \Gamma_0 + \sigma_{ac}T + \gamma_{LO}N_{LO}(T)$, where $\Gamma_0$ is a temperature-independent term due to scattering with impurities and defects, while the second and third terms correspond to homogeneous acoustic and LO phonon scatterings, with proportional factors $\sigma_{\mathrm{ac}}$ and $\gamma_{\mathrm{LO}}$, respectively. For LO phonons, the coupling strength is proportional to their population function $N_{\mathrm{LO}}(T)=1/[\exp\left(\frac{E_{\mathrm{LO}}}{k_{\mathrm{B}}T}\right)-1]$, where $k_{\mathrm{B}}$ is the Boltzmann constant and $E_{\mathrm{LO}}$ is the representative LO phonon energy suggested by the line broadening analysis. The fit yields $\Gamma_0$=11.4 meV, $\sigma_{\mathrm{ac}}$=0.053 meV $\mathrm{K}^{-1}$, $\gamma_{\mathrm{LO}}$=44.5 meV and $E_{\mathrm{LO}}$=15.7 meV. This LO phonon energy is consistent with other lead-halide perovskites,[24] and the dominant cause of homogeneous broadening in $CsPbBr_3$ is the Fröhlich interaction between excitons and LO phonons.

**Long-range transport probed by transient grating spectroscopy**

We proceeded to investigate the diffusive transport processes in $CsPbBr_3$ using TG spectroscopy. The schematic is illustrated in Figure 2a. Two noncolinear pump laser pulses, with wavevectors $\boldsymbol{k}_1$ and $\boldsymbol{k}_2$, interfere at the sample surface, creating a sinusoidal optical interference pattern that generates a grating of spatially modulated electron-hole pairs with a period $\Lambda$. The reciprocal space grating vector $\boldsymbol{q}$, with magnitude $q=2\pi/\Lambda$, is the difference between the two pump wavevectors $\boldsymbol{k}_1$ and $\boldsymbol{k}_2$, and that can be conveniently varied in our experiments by changing the phase mask array (Figure S6). The photoinduced electron-hole density wave modulates the local refractive index, acting as a "transient grating". A time-delayed probe laser pulse, with wavevector $\boldsymbol{k}_3$, is then diffracted by this grating to the direction $\boldsymbol{k}_3+\boldsymbol{q}$ and mixed with another probe pulse (the local oscillator), with wavevector $\boldsymbol{k}_4$, for heterodyne detection (see Methods for details). The time evolution of the grating modulation depth encodes information about both local population recombination and lateral diffusion (Figure 2b), which is tracked by recording the temporal decay of the diffracted TG signal.

In the TG experiment, we probed long-range exciton diffusion within a large (hundreds of μm) single-domain of $CsPbBr_3$ nanosheets on a sapphire substrate (Figure S2). Figures 2c and d present the TG signals as a function of time after photoexcitation, shown in semi-logarithmic plots for varying grating periods at cryogenic and room temperatures, respectively. At very low temperatures ($T < 50$ K), the TG signal exhibits a biexponential decay, as exampled in Figure 2c. The weak fast component, corresponding to the cooling of hot carriers,[50] has a grating independent lifetime of several ps (Figure S9). In contrast, the dominant slow component, with a decay time of several hundred ps, strongly depends on the grating period, indicating diffusive transport. Above 50 K, the fast component disappears entirely, resulting in a purely mono-exponential decay that persists up to room temperature (Figure 2d).

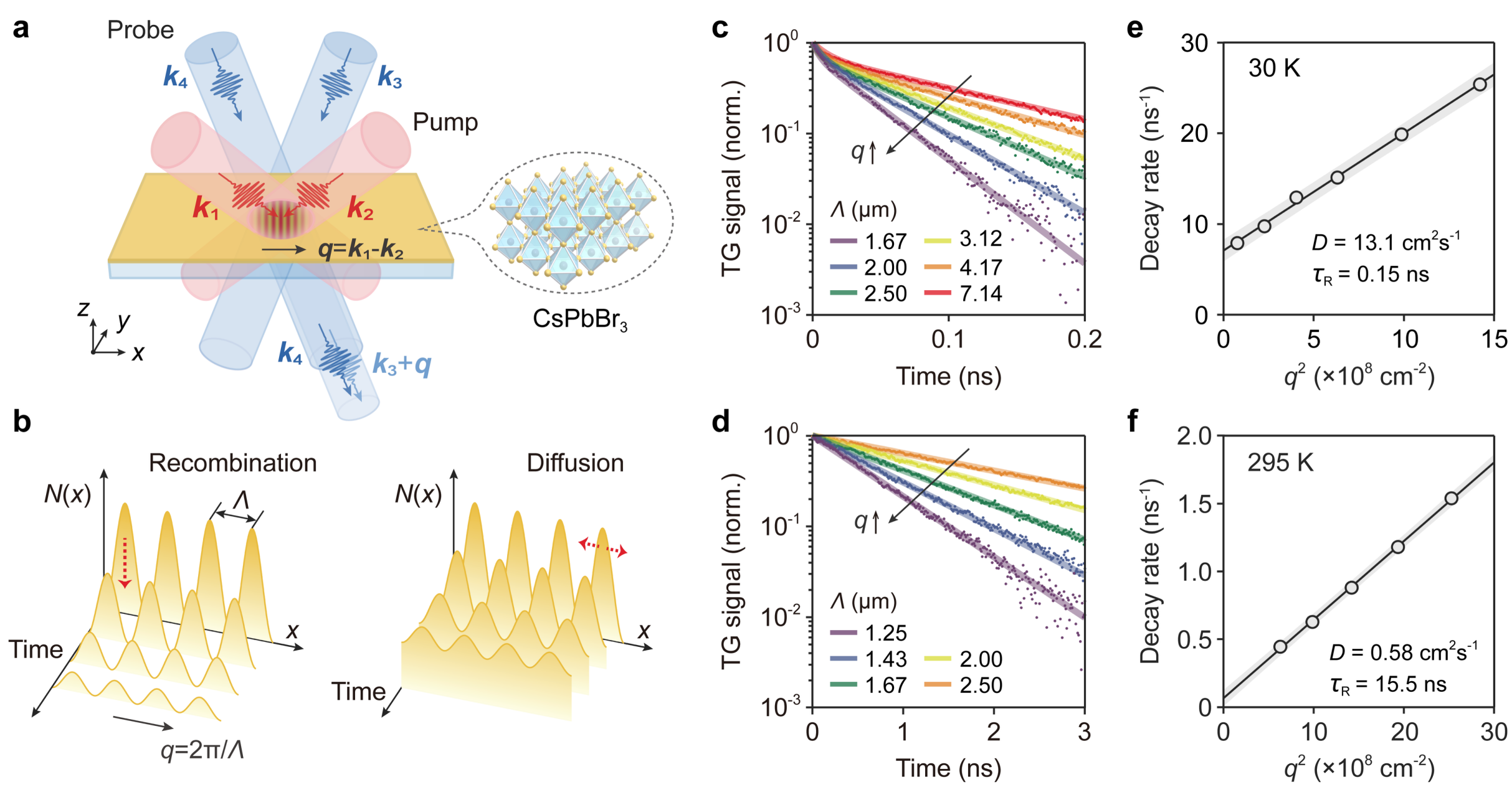

Figure 2. Transient grating spectroscopy for in-plane transport studies. (a) Schematic of the heterodyne transient grating (TG) technique. The "transient grating" is generated by the interference of two pump pulses and is detected *via* the diffraction of a probe pulse, which is mixed with another probe pulse at the detector for heterodyne detection. (b) Time evolution of the concentration profile $N(x)$ of the excited one-dimensional gratings through local population recombination (left) and lateral diffusion (right). (c and d) Normalized TG signal as a function of time for several grating wavevectors ($q$) measured at 30 K (c) and room temperature (d). Markers represent experimental data, and solid lines are exponential fits. (e and f) Grating decay rate ($\gamma$) versus $q^2$ at 30 K (e) and room temperature (f). The decay rate follows $\gamma = 1/\tau_R + Dq^2$, where $D$ is the exciton diffusion coefficient (obtained from the slope of the linear fit, solid black line), and $\tau_R$ is the exciton recombination lifetime (obtained from the inverse of the intercept). The shaded gray area indicates the 95% prediction intervals. Error bars are smaller than the marker sizes.

The TG signal decays faster with shorter grating periods due to the steeper concentration gradient.

As carriers diffuse, they move down the gradient until reaching an even distribution. Reducing the grating period (increasing $q$) intensifies the gradient, thus accelerating the diffusion process. Figures 2e and f show the extracted decay rate ($\gamma$) plotted as a function of $q^2$. The linear relationship with an intercept reflects the combined effects of exciton recombination and diffusion that follow $\gamma = 1/\tau_\mathrm{R} + Dq^2$, where $D$ is the diffusion coefficient for one-dimensional exciton transport along the grating wavevector direction and $\tau_\mathrm{R}$ is the lifetime associated with local exciton recombination. A linear fit yields the intercept as $1/\tau_\mathrm{R}$, while the slope directly gives the diffusion coefficient.

At 30 K, we obtained a large exciton diffusion coefficient of 13.1 $cm^2s^{-1}$ and a short recombination lifetime of 0.15 ns. At room temperature, the diffusion coefficient drops to a modest value of 0.58 $cm^2s^{-1}$, while the recombination lifetime extends to 15.5 ns. Combining the measured diffusion coefficient and lifetime, the exciton diffusion length can be calculated by $L_D = \sqrt{D\tau_R}$, which reaches ~1 μm at low optical excitation intensities at room temperature. Our TG measurements provide high-quality experimental data, confirmed by the consistency of values obtained from several samples and under varying excitation intensities (Figure S8).

### Determination of dominate scattering mechanisms in $CsPbBr_3$

We conducted temperature-dependent TG measurements from around 26 K to room temperature (Figure S7) to explore microscopic scattering mechanisms. The extracted diffusion coefficients are presented in Figure 3a as a function of temperature. As the temperature increases, no thermally activated increase in diffusion is observed, which points away from the hopping transport through defect-induced localized states. Instead, we observe a pronounced decrease in the diffusion coefficient within the first tens of K, followed by a gradual decline up to room temperature. On the contrary, the exciton recombination lifetime steadily increases from 0.1 ns at 26 K to 15.5 ns at room temperature (as discussed in detail in the next section). Due to the trade-off between diffusion and recombination, the corresponding diffusion length shows a slight increase with rising temperature, ranging from 0.4 to 1 μm (inset of Figure 3a).

To compare with charge transport studies, we converted the measured diffusion coefficient to an effective mobility using the Einstein relation $\mu = eD/(k_B T)$, where $e$ is the elementary charge. The effective exciton mobility is plotted as a function of temperature in Figure 3b. Similar to the diffusivity, the mobility decreases consistently with increasing temperature, from approximately $10^4$ $cm^2V^{-1}s^{-1}$ at 26 K to 22.8 $cm^2V^{-1}s^{-1}$ at room temperature, indicating phonon-limited band-like transport ($d\mu/dT < 0$) over the entire temperature range. In contrast, impurity scattering, commonly observed in unintentionally doped semiconductors at low temperatures,[27] leads to a thermally activated hopping transport ($d\mu/dT > 0$).

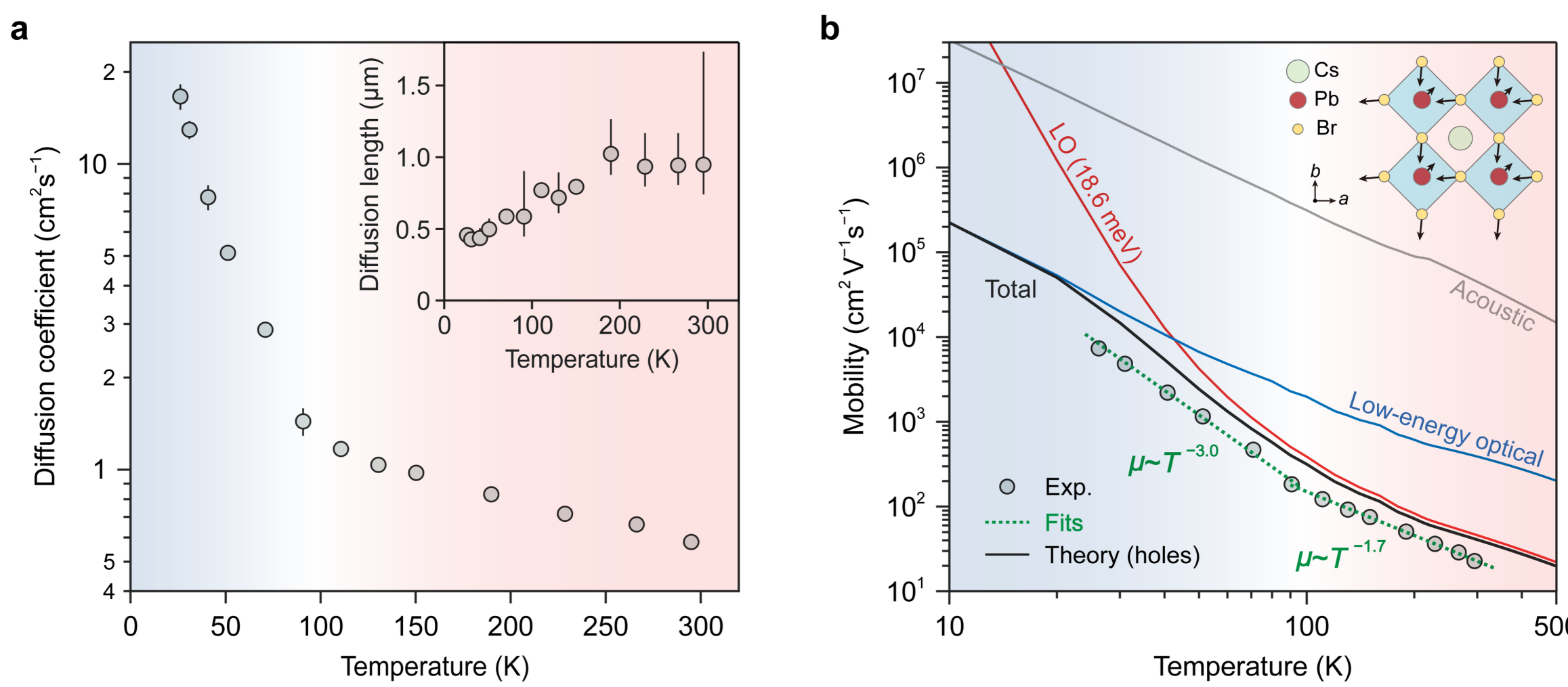

Figure 3. Temperature-dependent exciton transport. (a) Temperature-dependent diffusion coefficients obtained from TG experiments. The inset shows the diffusion length ($L_D$) calculated from the measured diffusion coefficient ($D$) and recombination lifetime ($\tau_R$) through the relation $L_D = \sqrt{D\tau_R}$. (b) Temperature-dependent mobilities. The gray dots represent exciton effective mobilities, converted from diffusion coefficients with the Einstein relation, and the dashed green lines show power-law fits to the experimental data in two distinct temperature ranges. By employing *ab initio* Boltzmann transport calculations, the black line gives the total phonon-limited mobilities for $CsPbBr_3$. Contributions from the highest energy (18.6 meV) LO phonon mode (red line), all other optical phonons (blue line), and acoustic phonons (gray line) are also displayed. Acoustic phonons clearly have a negligible impact on mobility over the entire temperature range. Instead, carrier transport is determined by low-energy optical phonons at low temperatures and by the highest energy LO phonon at high temperatures. The inset illustrates the atomic displacements (Pb-Br stretching) associated with the specific LO phonon mode.

During the transport of carriers, scattering events alter their momentum, thereby reducing their mobility. Mobility typically exhibits a power-law dependence on temperature, expressed as $\mu \sim T^{\alpha}$. The exponent $\alpha$, derived from empirical fits in prior studies of conventional semiconductors, can provide qualitative information about the dominate scattering mechanisms within specific temperature ranges. Note that although excitons are electrically neutral, their motion is still influenced by scattering events with phonons or impurities, and these interactions also impact exciton transport and provide valuable insights into the material's defect density and electron-phonon coupling.

For the experimentally extracted mobility data (gray dots in Figure 3b), we observe a rapid decrease at low temperatures, following a power law of $\mu \sim T^{-3.0}$. Above a turning point around 100 K, the mobility continues to decrease but at a slower rate, best fitted by $\mu \sim T^{-1.7}$. Although the high-temperature trend and values match well with previous electrical studies on similarly prepared $CsPbBr_3$ single crystals,[11,12] the rapid decrease below 100 K is observed here for the first time. These scaling behaviors significantly deviate from conventional expectations, where gapless acoustic phonons are predicted to dominate at low temperatures, exhibiting a $T^{-1.5}$ dependence, while optical phonons are expected to accelerate the decline at higher temperatures.

To resolve the discrepancy, we performed advanced first-principles calculations of phonon-limited carrier mobilities within the framework of the Boltzmann transport equation (see Note S3 for details). The calculated trends, accounting for scattering by different phonon modes, are shown in Figure 3b. The ultimate phonon-limited mobility (black line) from the calculations closely aligns with the experimental data (gray dots), though the calculated values are slightly higher. Here we present hole mobility as a representative case, with electron mobility exhibiting similar behavior (Figure S11). Note that exciton mobility cannot be simply derived from the sum of the electron and hole scattering rates, as suggested by Matthiessen's rule, nor by the effective mass approximation. The scattering rate of excitons depends not only on the individual cases of electrons and holes but also on the interplay between their electronic clouds, which modifies long-range electrostatic interactions with phonons. We anticipate that the temperature dependence of exciton mobility will follow a trend similar to that of hole (or electron) mobility, with the exact values differing slightly. A rigorous approach requires direct calculations of the exciton-phonon matrix elements and relaxation lifetimes, which treats the exciton as a combined entity[51] and is beyond the scope of this study. The perfectly aligned trends indicate that the essential physics of the system is effectively captured in the calculations.

Now, we turn to analyzing the mechanisms that dominate transport across various temperature ranges. As shown by the red line in Figure 3b, a single LO phonon with the highest energy (18.6 meV) governs room-temperature mobility, nearly coinciding with the total mobility (black line) above 100 K. This is not unexpected, as LO phonons induce stronger scattering through long-range Coulomb interactions, driven by microscopic electric fields resulting from polar atomic displacements (see inset in Figure 3b). The dominant role of this 18.6-meV LO phonon mode in electron-phonon coupling is further supported by recent resonant Raman experiments[52] and transport calculations for $CsPbBr_3$.[53]

At temperatures below ~30 K, the mobility is limited primarily by scattering from low-energy optical phonons (excluding the highest-energy LO mode) rather than from acoustic phonons. Due to the quasi-continuous distribution of these low-energy optical modes near the Γ point in the phonon dispersion (Figure S10), their cumulative scattering effect (blue line) mimics that of acoustic phonons (gray line) but with greater strength. In the intermediate temperature range (30-100 K), the measured scaling behavior of $\mu \sim T^{-3.0}$ arises from combined scattering by low-energy optical phonons and the dominant LO phonon.

Notably, the switch between the two scaling laws in our sample is unrelated to the structural phase transition, which occurs at a much higher temperature of 260 K. Despite the reduction in crystal symmetry from cubic to orthorhombic, the phonon modes and energies remain largely similar, as indicated by Raman spectra (Figure 1c) and phonon dispersions (Figure S10), with the LO phonon of similar energy continuing to dominate the electron-phonon coupling strengths. As a result, we expect the mobility to exhibit nearly identical temperature trends across the different phases, with only slight variations validated as negligible in experiments. The sudden change observed in $MAPbBr_3$ is likely related to the freedom of MA cation rotation associated with the phase transition.[10,54] Moreover, no distinction in mobility was observed along different in-plane directions of the sample, which could be rationalized by the isotropic LO phonon energies along different high-symmetry lines in the phonon dispersions (Figure S10).

**Intrinsic exciton radiative recombination**

In addition to the mobility results, the recombination lifetime is found to increase monotonically with temperature, a trend that is typically associated with intrinsic radiative recombination. To further investigate this behavior, we carried out time-resolved photoluminescence (TRPL) spectroscopy to examine the effective decay dynamics of radiative species. The excitation is non-resonant and slightly above the band gap, with an ultralow excitation power corresponding to a carrier density of ~$10^{15}$ $cm^{-3}$ (Note S2).

Figure 4a displays the PL transients of the $CsPbBr_3$ excitonic emission at various temperatures. The effective decay time, extracted from a mono-exponential fit of the measured kinetics, is plotted as a function of temperature in Figure 4b (red squares). For comparison, recombination lifetimes extracted from the TG experiment (gray dots) are also included, exhibiting a similar temperature dependence. The relatively long PL decay times may be related to slight deviations from mono-exponential decay dynamics, as well as differences in the temporal windows between the two techniques.[55] Notably, at elevated temperatures, the PL transients exhibit more pronounced deviations from a perfect exponential shape, likely due to enhanced contributions from various nonradiative processes. These may include thermally activated trap-assisted recombination, optically forbidden transitions, and thermal ionization of excitons.

Most previous TRPL studies have reported a decrease in lifetime with increasing temperature in halide perovskites, likely due to thermally activated non-radiative recombination caused by defects. In contrast, our observations reveal intrinsic radiative behavior in high-quality $CsPbBr_3$, where the contribution of defect-related processes is negligible, particularly at low temperatures. Radiative recombination can occur through either excitons or free carriers; however, the nearly single-exponential kinetics observed in both TG and TRPL experiments suggest a geminate exciton recombination, confirmed by its timescale on the order of ns.[56]

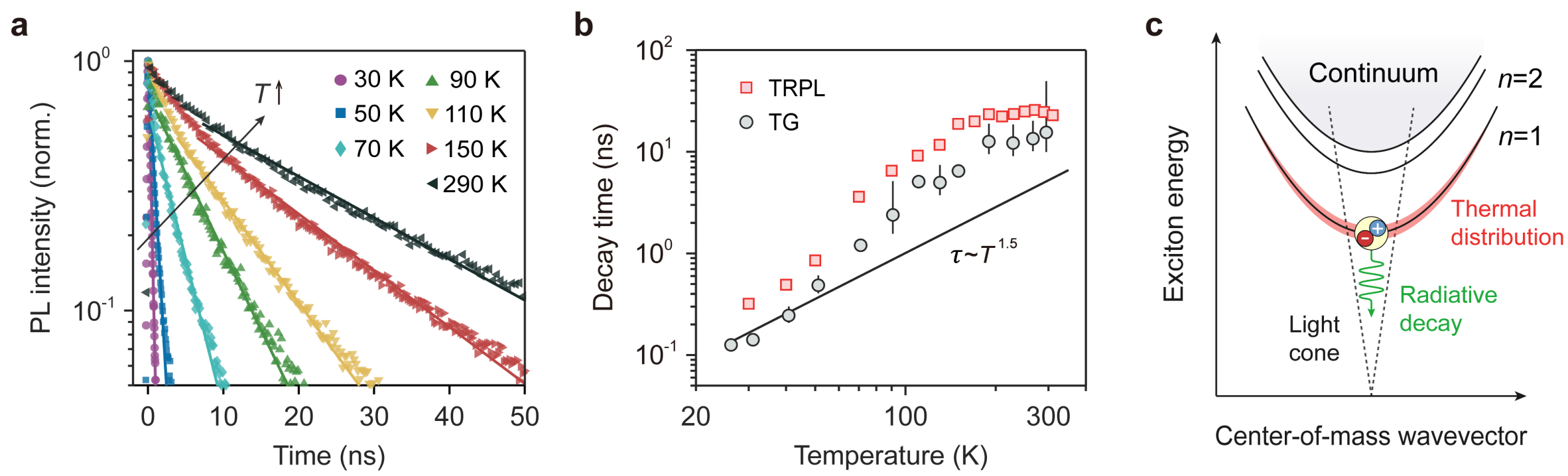

Figure 4. Temperature dependence of PL decay dynamics. (a) Normalized TRPL decays measured at various temperatures. Markers represent experimental data, and solid lines correspond to mono-exponential fits. (b) PL decay times as a function of temperature, shown together with recombination lifetimes extracted from TG measurements for comparison. Both exhibit consistent increasing trends, though notable deviations arise due to different observation timescales. The black line represents the intrinsic temperature dependence of the exciton radiative lifetime ($\tau \sim T^{1.5}$), as predicted by an empirical model that considers the thermal population of exciton states. (**c**) Momentum conservation and radiative decay. After photoexcitation, excitons rapidly establish a thermal distribution (shaded red area). Momentum conservation imposes a restriction on radiative decay, allowing only for excitons with near-zero wavevector, confined within the light cone (dashed lines) defined by the linear dispersion relation of light in the crystal.

Furthermore, the temperature trend and absolute values observed in Figure 4b closely resemble those reported for $CsPbBr_3$ nanocrystals,[33] supporting the conclusion that the dominant species is indeed excitonic in nature. However, the photoluminescence quantum yield (PLQY) at room temperature remains below 5% for our samples, which suggests that the overall radiative efficiency is limited. Compared to single crystals, $CsPbBr_3$ nanocrystals exhibit significantly higher PL efficiencies due to intrinsic quantum confinement. With proper surface treatment and passivation, their PLQY can even exceed 90%.[57] In contrast, the relatively low values commonly measured in bulk single crystals[58] may result from competitive non-radiative recombination at room temperature, or from extrinsic factors such as incomplete photon collection and photon reabsorption within the crystal.

At the lowest temperature in our experiments, recombination occurs very rapidly, approaching the temporal resolution of our TRPL system (~200 ps, Figure S12). This timescale is comparable to the intrinsic exciton lifetime of 125 ps reported for high-purity GaN at 2 K.[59] The rapid and efficient luminescence observed in our $CsPbBr_3$ sample is a signature of intrinsic exciton radiative recombination, commonly observed in semiconductors with large exciton binding energies.[31,32] The ultrashort radiative lifetime exhibited by $CsPbBr_3$ highlights its potential as a strong radiative emitter, particular in nanocrystalline forms[30].

The temperature dependence of the decay time can be explained by momentum conservation in the radiative recombination process.[31] In this process, only bright excitons with near-zero center-of-mass momentum (within the light cone) can couple with light and recombine radiatively. Since light possesses a very small wavevector (~$10^5$ $cm^{-1}$), the light cone is narrowly defined by the light dispersion curves in the crystal, as illustrated in Figure 4c. At very low temperatures and under weak excitation, most excitons remain coherent within the light cone, efficiently couple with light and decay rapidly, resulting in an ultrashort radiative lifetime. However, as the temperature rises, excitons are increasingly scattered by phonons into high-momentum states, forming a thermal equilibrium distribution (shaded red area in Figure 4c) on the ps timescale. Excitons outside the light cone occupy non-radiative, momentum-dark states that must relax back into the light cone to undergo radiative recombination, leading to a longer exciton radiative lifetime with increasing temperature. This reduction in radiative recombination rate also contributes to the decrease in PLQY with increasing temperature.[33]

For a parabolic band and a simple Boltzmann distribution, the exciton radiative lifetime $\tau$ follows the well-known temperature dependence $\tau \sim T^{1.5}$ in bulk semiconductors (see detailed derivation in Note S4). However, at elevated temperatures, the experimental trend (Figure 4b) deviates from this predicted behavior, likely due to the gradual dissociation of excitons into free electrons and holes.[32,60] Although the radiative lifetime is expected to scale as $T^{1.5}$ for both excitons and free carriers, the latter typically results in a much lower recombination rate.

The competition with free-electron-hole recombination is inevitable for bulk crystals with a relatively modest exciton binding energy, leading to a significant increase in the average lifetime. The further prolongation of the radiative lifetime makes non-radiative channels more competitive, which may partly explain why the PLQY remains low even in very high-quality single crystals. Similar trends have been observed in materials like GaN,[32] where the radiative lifetime increases more rapidly than expected at relatively high temperatures. Despite these experimental observations on coupled exciton-carrier dynamics, interpreting precise radiative lifetimes through first-principles calculations still requires further development.[61]

## DISCUSSION

Our study reveals intrinsic exciton transport and recombination behaviors over the entire temperature range in $CsPbBr_3$ single crystals, enabled by reliable, contact-free measurements using transient grating spectroscopy. Prior to our study, this technique had been applied to poly-crystalline samples, where grain boundaries acted as hard barriers and severely hindered carrier diffusion.[62] Hence, we emphasize the critical role of epitaxial growth in preparing large-area, single-crystalline thin films to facilitate the study of intrinsic long-range transport. The effective exciton mobility decreases continuously from 26 K to room temperature due to enhanced phonon scattering, whereas the lifetime shows the opposite trend, increasing steadily as a result of thermal quenching of exciton radiative recombination.

Among the various phonon scattering mechanisms, LO phonon scattering has the greatest impact due to the strong Coulomb interaction. The thermal activation of LO phonon scattering typically causes a drastic reduction in carrier mobility at elevated temperatures and becomes the limiting factor near room temperature in many traditional polar semiconductors. Metal halide perovskites differ from these more rigid compounds through their uniquely soft lattice structure, resulting in low-energy LO phonon modes, typically less than 20 meV. For example, in $CsPbBr_3$, DFT calculations predict the highest-energy LO phonon mode to be 18.6 meV, which agrees well with the 15.7 meV value obtained from line broadening analysis (Figure 1e). For comparison, the LO phonon energy is 36 meV in GaAs[63] and reaches up to 92 meV in GaN[64]. Therefore, the rapid decline in mobility observed at fairly low temperatures (Figure 3b) in $CsPbBr_3$ is attributed to the onset of soft LO phonon scattering. Once LO phonon scattering fully dominates, the rate of mobility decline slows down. The observed $\mu \sim T^{-1.5}$ dependence at room temperature appears to be coincidental and is expected to slow further as the temperature continues to increase. We show that in perovskite systems, where many phonon modes with comparable energies contribute to scattering simultaneously, the direct application of empirical theories can be highly inaccurate, particularly when incorporating the activation of LO phonon scattering.

It is surprising that defect-related hopping transport remains negligible even at the lowest temperature in our study. Although epitaxial growth may facilitate the observation of intrinsic behaviors by reducing defect density, the 'defect-free' characteristic of perovskites is primarily attributed to their exceptional defect tolerance, which arises from the formation of large polarons and dynamically screened Coulomb potential.[21] A recent study also indicated that coupling with the low-frequency lattice phonons could explain the defect tolerance particularly toward radiative recombination.[65] Therefore, the high defect tolerance seems to be closely correlated with the soft nature of halide perovskites as well.

Another issue that needs further discussion is the excitons versus free carriers. The relatively large $E_b$ of 35 meV, obtained from Rydberg exciton states in absorbance spectra, suggests that excitons can resist thermal ionization at room temperature, which is supported by the observation of exciton polariton condensation in epitaxial $CsPbBr_3$.[29] We note that the Saha equation predicts an exciton population of only a few percent at room temperature when $E_{\mathrm{b}}$ is less than 50 meV.[66] However, the dominance of free carriers conflicts with the single-exponential decay and recombination timescales in our time-resolved measurements. Recent studies have pointed out that non-equilibrium recombination is governed by kinetics rather than thermodynamics, and the Saha equation may significantly overestimate the role of free carriers in recombination.[67] Therefore, we consider excitons to be the dominant recombination species, though the average decay rate is influenced by the coexistence of free carriers.

In summary, we demonstrate intrinsic transport behaviors in high-purity $CsPbBr_3$ down to 26 K and clearly resolve the underlying mechanisms through combined experimental and calculational studies. The rapid decline in mobility with $\mu \sim T^{-3.0}$ below 100 K is measured for the first time, identifying the activation of strong polar optical phonon scattering. The well-known $\mu \sim T^{-1.5}$ trend above 100 K can be unambiguously attributed to the dominant scattering from single LO phonons, rather than from acoustic phonons or polarons. The softness of halide perovskites gives rise to a series of low-energy optical phonons, distinguishing their behaviors from rigid traditional semiconductors. We also detect the rarely observed intrinsic radiative recombination, with excitonic emission showing much higher recombination rates and efficiency, highlighting the potential of $CsPbBr_3$ for efficient light-emitting applications. Importantly, many of the conclusions drawn here are also applicable to nanoscale systems. While exciton transport in nanostructures is often complicated by grain boundaries and interfacial effects, recombination processes could be simplified due to spatial confinement and the suppression of non-radiative channels. Our work contributes to a deeper understanding of the intrinsic physics of metal halide perovskites, paving the way for future modifications in their electrical and optical properties.

## METHODS

**Growth and Transfer of Single-Crystalline $CsPbBr_3$.** Large-area single-crystalline $CsPbBr_3$ thin films were grown using the chemical vapor transport method with a custom-built chemical vapor deposition (CVD) system. High-purity cesium bromide (CsBr) and lead bromide ($PbBr_2$) precursors were thoroughly mixed in a 1:1 molar ratio as the source material. Muscovite mica, used as the substrate, was cleaned with acetone and placed downstream in a quartz tube mounted in a tube furnace. Prior to growth, the tube was evaluated and purged several times with high-purity nitrogen to ensure an inert atmosphere. During the growth process, crystalline $CsPbBr_3$ was epitaxially grown on the mica substrate. The temperature and pressure inside the quartz tube were set to 575 °C and 20 mTorr, respectively, and stabilized for 5 min. After growth, the tube was cooled down naturally. The resulting perovskite thin films can be mechanically lifted off and transferred to a target substrate (such as sapphire) using a dry transfer method with Scotch tape. Although bulk $CsPbBr_3$ typically adopts an orthorhombic structure at room temperature,[42] the as-grown thin films (several hundred nm thick) exhibit a tetragonal phase. This phase shift is likely due to the combined effects of the high synthesis temperature and significant surface energy contributions.[68]

**XRD Measurements.** Powder XRD measurements were performed using a Rigaku SmartLab diffractometer with Cu Kα radiation in Bragg-Brentano geometry at room temperature. Structural properties and phase transitions were further investigated through single-crystal XRD on an XtaLAB Synergy diffractor equipped with a liquid nitrogen cryostat. For single-crystal measurements, free-standing $CsPbBr_3$ samples were prepared to eliminate strong diffraction signals from the sapphire substrate. Laue diffraction patterns were recorded over a temperature range of 80-400 K. Structure refinement revealed a cubic phase above 300 K, an orthorhombic phase below 260 K, and a tetragonal phase in the intermediate range.

**Raman Measurements.** Polarized Raman spectra were measured in the backscattering

configuration over a temperature range from 4 to 300 K. A 633-nm continuous-wave HeNe laser was used as the excitation source. A microscope objective (numerical aperature=0.82) focused the incident laser into a ~2 μm spot, and the scattered Raman signal was collected by the same objective. The spectra resolution was approximately 0.2 $cm^{-1}$ per CCD pixel, achieved with a 2400 lines/mm grating. The unpolarized spectra presented in the main text were obtained by averaging over all polarizations.

**Temperature-Dependent Absorbance and PL Measurements.** The sample was mounted on the cold finger of a closed-loop helium cryostat under vacuum, allowing for temperature variation from 10 to 340 K. For absorbance experiments, a stabilized quartz-tungsten lamp served as the white light source. The beam was efficiently focused and coupled into a single-mode fiber with a core diameter of 2.5 μm. Light emitted from the fiber, acting as a practical point source, was recollimated and refocused onto the sample *via* an objective, creating a ~5 μm diameter spot. The transmitted light was collected by a multi-mode fiber coupled to the entrance slit of a Czerny-Turner spectrometer. The transmitted spectra of both the sample and the pure substrate were recorded separately by a thermoelectrically cooled EMCCD to calculate absorbance. For PL experiments, the sample was excited at oblique incidence by a 405-nm continuous-wave diode laser under low excitation conditions. The emission was collected in the normal direction and filtered through a 450-nm long-pass filter.

**Transient Grating Spectroscopy.** In the transient grating experiment (schematically illustrated in Figure S6), 200 fs laser pulses were generated by an 80-MHz mode-locked Ti:Sapphire laser combined with an optical parametric oscillator (OPO) with an intra-cavity second-harmonic generator (SHG) to double the OPO signal frequency. The output laser was split into the pump and the probe beams at a degenerate wavelength of 526 nm (resonant band gap excitation) to maximize the signal. The pump beam intensity was modulated at 100 kHz by a photoelastic modulator combined with a linear polarizer. The probe beam was delayed by up to 13 ns relative to the pump beam using an optical delay stage.

Both the pump and probe beams were then aligned parallel with a horizontal offset and focused onto a custom-made transmission grating, referred to as the "phase mask", using a lens. The phase mask diffracted the beams into multiple vertical orders. The primary ±1 diffraction orders were isolated to form two coherent pump beams and two time-delayed coherent probe beams. These four beams, arranged in a boxcar phase-matching geometry,[69] were focused onto the same spot of the sample using a large spherical mirror. Beam overlap was optimized with a pinhole, and the spot sizes were estimated to be ~150 μm for both pump and probe beams. A 3D translation stage mounted with an array of phase masks with varying periods ($d$) allowed for fast switching of the grating period.

The two coherent pump beams interfered on the sample surface, generating an electron-hole grating with a period of $\Lambda = d/2$. A probe beam was then diffracted by the grating and mixed with the other probe beam, which served as a local oscillator for heterodyne detection.[69] To modulate the phase difference between the diffracted probe and the local oscillator, a thin coverslip (~170 μm thick) was mounted on a galvanometric scanner operated at 250 Hz. The dynamics of the transient grating was monitored by demodulating changes in the diffracted probe signal as a function of time delay using two lock-in amplifiers. The sample was mounted on the cold finger of a closed-loop helium cryostat under vacuum, enabling temperature-dependent measurements.

**Time-Resolved PL Spectroscopy.** Time-resolved PL was performed by exciting the sample with ~200 fs pulses at 430 nm, generated by frequency doubling of an 80-MHz mode-locked Ti:Sapphire laser. The pulse interval was extended to 0.5 μs using an acousto-optical pulse selector with a 40:1 division ratio. Similar to the steady-state PL setup, the laser was focused to a spot ~10 μm in diameter at oblique incidence, and the emitted PL was collected in the normal direction, coupled into a multi-mode fiber, and delivered to the entrance slit of a Czerny-Turner spectrometer. After dispersion, a specific wavelength was selected by the exit slit, and the photons were focused onto a single-photon avalanche photodiode. The temporal decay of the PL was measured using a time-correlated single-photon counter, with the detector's counting rate controlled at ~1% to avoid pile-up effects. The overall time resolution of the system was determined to be ~200 ps (Figure S12). All measurements were conducted at extremely low laser fluences, corresponding to a typical photogenerated carrier density of ~$10^{15}$ $cm^{-3}$ (see Note S2 for detailed calculations).

**First-Principles Calculations.** Density functional theory (DFT) calculations were performed using the Perdew-Burke-Ernzerhof (PBE) exchange-correlation functionals in the Quantum ESPRESSO package.[70] All quantities, including the electronic structure, phonon dispersion and electron-phonon coupling matrix, were derived from first principles. To simplify the computations, the high-symmetry cubic phase of $CsPbBr_3$ was applied throughout the calculations. The plane-wave energy cutoff was set to 100 Ry, and norm-conserving pseudopotentials were obtained from the SG15 repository.[71] In the cubic phase, the lowest-energy phonons exhibited imaginary mode frequencies within the harmonic approximation, reflecting the dynamically unstable nature of the high-temperature phase.[72] To address this, we employed ZG.x[73] to obtain a stable, renormalized phonon dispersion at 300 K on a 2×2×2 supercell (Figure S10).

The calculations focused on phonon-limited transport properties, given the band transport observed in the experiment. The phonon-limited mobility was calculated using the Boltzmann transport equation (BTE) implemented in the EPW code.[74] First, the electron-phonon coupling matrix elements were computed on a coarse grid, and subsequently interpolated onto a fine grid to ensure convergence in mobility calculations. Electron and hole mobilities were computed using both the self-energy relaxation time (SERTA) approximation and the iterative BTE (IBTE) approach (Note S3) with a carrier density of $1\times10^{18}$ $cm^{-3}$ (see Figure S11). A 120×120×120 mesh grid was employed for fine-grid interpolation of the electron-phonon coupling matrix elements to compute mobilities.

To simulate temperature dependence, slight changes in the band structure and phonon frequencies due to thermal expansion were neglected. Instead, temperature effects were incorporated through the thermal equilibrium distribution of phonons. Additionally, the isotropic Eliashberg spectral function was presented to show the distribution of electron-phonon coupling strengths across different phonon modes and evaluate their relative contributions to transport (see Figure S10).

## ASSOCIATED CONTENT

**Supporting Information**. Additional experimental details and analysis, including notes on the Elliot model (Note S1), excitation density calculation (Note S2), Boltzmann transport equation (Note S3), and exciton radiative lifetime (Note S4). Supplementary Figures include crystal structures and single-crystal XRD patterns at different phases (Figure S1); optical microscopy image of dispersed thin films transferred onto a sapphire substrate (Figure S2); AFM image (Figure S3); exciton Rydberg states and Elliot fit (Figure S4); temperature-dependent absorbance and PL (Figure S5); TG setup (Figure S6); temperature-dependent TG measurements (Figure S7); the effect of probe wavelength and excitation power (Figure S8); fast component lifetime at 30 K (Figure S9); phonon dispersions and electron-phonon coupling strength (Figure S10); separate electron and hole transport (Figure S11); and instrument response function for TRPL experiments (Figure S12).

## AUTHOR INFORMATION

**Corresponding Authors**

**Sheng Meng** − Beijing National Laboratory for Condensed Matter Physics and Institute of Physics, Chinese Academy of Sciences, Beijing, 100190, China; Email: smeng@iphy.ac.cn

**Qihua Xiong** − State Key Laboratory of Low Dimensional Quantum Physics, Department of Physics, Tsinghua University, Beijing 100084, China; Email: qihua_xiong@tsinghua.edu.cn

**Luyi Yang** − State Key Laboratory of Low Dimensional Quantum Physics, Department of Physics, Tsinghua University, Beijing 100084, China; Email: luyi-yang@mail.tsinghua.edu.cn

## Author Contributions

Z.B. and L.Y. conceptualized the project. Y.S. and Q.X. prepared the samples and performed the AFM measurements. T.S. and Z.B. performed the XRD measurements. H.W., Z.B., M.L., and P.T. performed the Raman measurements. Z.B. and L.Y. performed the absorbance and PL measurements. Z.B., H.Z., and L.Y. performed the TG measurements. Z.B., Y.W., Q.X., and L.Y. performed the TRPL measurements. Y.Z., D.Z., and D.M. performed PLQY measurements. Y.B., Y.W., and S.M. performed the first-principles calculations. Z.B., Y.B., and L.Y. analyzed the data. Z.B., Y.B., Y.S., S.M., Q.X., and L.Y. contributed to the discussion. L.Y. supervised the work. All authors contributed to the writing of the manuscript.

## Funding Sources

This work was supported by the National Natural Science Foundation of China (Grants No. 12361141826, No. 12421004, and No. 12074212), National Key R&D Program of China (Grants No. 2021YFA1400100 and No. 2020YFA0308800), Natural Science Foundation of Beijing, China (Grants No. Z240006).

## Notes

The authors declare that they have no competing interest.

ACKNOWLEDGMENT

This work was supported through the Synergetic Extreme Condition User Facility (SECUF, https://cstr.cn/31123.02.SECUF).

ABBREVIATIONS

AFM, atomic force microscopy; BTE, Boltzmann transport equation; CVD, chemical vapor deposition; DFT, density functional theory; FWHM, full width at half maximum; IBTE, iterative Boltzmann transport equation; LO, longitudinal optical; PL, photoluminescence; PLQY, photoluminescence quantum yield; SERTA, self-energy relaxation time approximation; TG, transient grating; TRPL, time-resolved photoluminescence; XRD, X-ray diffraction.